\documentclass[aps,showpacs,onecolumn,superscriptaddress,preprint]{revtex4-1}

\usepackage{graphicx}
\usepackage{subfigure}
\usepackage{bm}
\usepackage{color}
\usepackage{array}
\usepackage{epstopdf}
\usepackage{soul}

\newcommand{\Ga}{\Gamma}

\begin{document}

\title{Projective Quasiparticle Interference of a Single Scatterer \\ to Analyze
the Electronic Band Structure of ZrSiS}

\author{Wenhao Zhang}
\affiliation{Zhejiang Province Key Laboratory of Quantum Technology and Device, Department of Physics, Zhejiang University, Hangzhou, 310027, China}
\author{Kunliang Bu}
\affiliation{Zhejiang Province Key Laboratory of Quantum Technology and Device, Department of Physics, Zhejiang University, Hangzhou, 310027, China}
\author{Fangzhou Ai}
\affiliation{Zhejiang Province Key Laboratory of Quantum Technology and Device, Department of Physics, Zhejiang University, Hangzhou, 310027, China}
\author{Zongxiu Wu}
\affiliation{Zhejiang Province Key Laboratory of Quantum Technology and Device, Department of Physics, Zhejiang University, Hangzhou, 310027, China}
\author{Ying Fei}
\affiliation{Zhejiang Province Key Laboratory of Quantum Technology and Device, Department of Physics, Zhejiang University, Hangzhou, 310027, China}
\author{Yuan Zheng}
\affiliation{Zhejiang Province Key Laboratory of Quantum Technology and Device, Department of Physics, Zhejiang University, Hangzhou, 310027, China}
\author{Jianhua Du}
\affiliation{Zhejiang Province Key Laboratory of Quantum Technology and Device, Department of Physics, Zhejiang University, Hangzhou, 310027, China}
\author{Minghu Fang}
\affiliation{Zhejiang Province Key Laboratory of Quantum Technology and Device, Department of Physics, Zhejiang University, Hangzhou, 310027, China}
\affiliation{Collaborative Innovation Center of Advanced Microstructures, Nanjing University, Nanjing 210093, China}
\author{Yi Yin}
\email{yiyin@zju.edu.cn}
\affiliation{Zhejiang Province Key Laboratory of Quantum Technology and Device, Department of Physics, Zhejiang University, Hangzhou, 310027, China}
\affiliation{Collaborative Innovation Center of Advanced Microstructures, Nanjing University, Nanjing 210093, China}

\begin{abstract}
Quasiparticle interference (QPI) of the electronic states has been widely applied in scanning tunneling microscopy (STM)
to analyze electronic band structure of materials. Single-defect-induced QPI reveals defect-dependent interaction
between a single atomic defect and electronic states, which deserves special attention. Due to the weak signal of
single-defect-induced QPI, the signal-to-noise ratio (SNR) is relatively low in a standard two-dimensional
QPI measurement. In this paper, we introduce a projective quasiparticle interference (PQPI) method, in which
a one-dimensional measurement is taken along high-symmetry directions centered on a specified defect.
We apply the PQPI method to a topological nodal-line semimetal ZrSiS.
We focus on two special types of atomic defects that scatter the surface and bulk electronic bands.
With enhanced SNR in PQPI, the energy dispersions are clearly resolved along with high symmetry directions.
We discuss the defect-dependent scattering of bulk bands with the non-symmorphic symmetry-enforced selection rules.
Furthermore, an energy shift of the surface floating band is observed and a new branch of energy dispersion (${\bm q}_6$)
is resolved. This PQPI method can be applied to other complex materials to explore defect-dependent interactions
in the future.

\end{abstract}

\maketitle

\section{Introduction}
\label{sec1}

In scanning tunneling microscopy (STM), quasiparticle interference (QPI) of
electronic states has been a powerful tool to analyze the electronic band structure of
condensed matter materials~\cite{Crommie1993,Hasegawa93, LCDavis91, Avouris94, Petersen98, Sprunger1764, PhHofmann97, Pascual04, Weismann1190,
Hoffman1148, Wang03, kostin2018imaging, Aynajian2012, Allan2013, Roushan2009, ZhangTong09, Zeljkovic2014, Chang2016, Inoue1184, Zheng2016, Batabyale1600709}.
QPI arises when the electronic state with initial momentum ${\bm k}_i$ is elastically
scattered to a final momentum ${\bm k}_f$. The potential barrier of
scattering is often induced by point defects, steps or other local perturbations
in materials. The scattering process leads to a spatial oscillation of electronic state
with wave vector ${\bm q}={\bm k}_f-{\bm k}_i$. The wave vector can be extracted from Fourier transform of QPI oscillations.
As a function of energy E, the ${\bm q}(E)$ dispersion reflects the electronic band structure in ${\bm k}$-space.

The QPI study initially focused on electronic surface state,
whose QPI oscillation (or Friedel oscillation) decays slowly with increasing distance
from the scattering center~\cite{Crommie1993, Hasegawa93, LCDavis91, Avouris94}.
QPI has been thereafter applied in both surface states of materials,
and electronic structure of two-dimensional (2D) materials~\cite{Petersen98, Sprunger1764, PhHofmann97, Pascual04}.
On the other hand, parallel features in Fermi surface structure may cause
anisotropic propagation of a three-dimensional (3D) band, which can also result in
a strong QPI oscillation on the sample surface~\cite{Weismann1190}.
The standard QPI measurement requires a 2D grid measurement, while in some special case
it can be reduced to $1D$ measurement due to a quasi $1D$ electronic structure near an
edge or homogeneous electronic structure induced by a single defect~\cite{Haim2018am,Liu_2017,Drozdov_2014}.
The development of QPI technique enables extensive analysis of band structure of complex
materials, including high-$T_\mathrm{c}$ superconductors~\cite{Hoffman1148, Wang03, kostin2018imaging},
heavy fermion systems~\cite{Aynajian2012, Allan2013},
and topological materials~\cite{Roushan2009, ZhangTong09, Zeljkovic2014, Chang2016, Inoue1184, Zheng2016, Batabyale1600709}.

Although all QPI oscillations are related to the underlying electronic
band structure, QPI induced by a single scatterer deserves special
attention~\cite{Inoue1184, Zheng2016, Simon2009, Derry2015}. Different
types of point defects trigger defect-dependent interaction between
the defect and electronic states. The QPI analysis around specified
point defects can reveal a selective scattering. For example, in topological
nodal-line materials ZrSiS and ZrSiSe~\cite{xu2015two, hu2016evidence, schoop2016dirac},
two different types of point defects are found to scatter electronic states of
the floating surface band~\cite{ZZ2018NatCom, Surface_floating} and bulk band,
respectively~\cite{lodge2017observation,butler2017quasiparticle,  Bu2018PRB, ZZ2018NatCom, surface_termination}.
In ZrSiSe, both the surface and bulk bands were observed to be scattered by
a single defect~\cite{Bu2018PRB, ZZ2018NatCom}, which has not been
detected yet in ZrSiS.

However, the QPI signal around a single scatterer is relatively weak, resulting in a low
signal-to-noise ratio (SNR) in the Fourier-transformed QPI pattern. The vague band
structure in this analysis limits the data-based discussion of physical properties.
Here in this paper, a new type of point defect is discovered in ZrSiS, and
a projective QPI (PQPI) method is introduced to analyze the scattered electronic
bands in two different point defects, with a much higher SNR. The first
new type of Zr-site defect scatters both the surface and bulk band. The second
type S-site defect only scatters the bulk band, which has been observed
before~\cite{lodge2017observation} with a different interpretation. A
preliminary 2D QPI measurement shows that the QPI pattern induced by
a single defect is anisotropic and highly concentrated along high symmetry
directions. With the PQPI method, we could clearly resolve the dispersion
branches and compare them with the density functional theory (DFT)
calculation. We discuss the selective scattering with non-symmorphic symmetry-related selection rules. We also observe a possible defect-induced energy shift of the
floating surface band, and an extra bulk band dispersion of ${\bm q}_6$ branch.
PQPI is a simple and intuitive method that can be applied in general
single scatterer induced QPI studies of different materials.

\section{Experimental Method}
\label{sec2}

Single crystals of ZrSiS were grown by a two-step chemical vapor transport method using iodine
as a transport agent~\cite{sankar2017crystal}. In the first step, a stoichiometric amount of 99.9 \%
purity precursors of Zr:Si:S = 1:1:1 molar ratio was pressed into a tablet, and put in an alumina crucible.
After sealed in an evacuated quartz ampoule, the sample was treated at 1100 $^\circ$C for two
days and then furnace-cooled to room temperature. In the second step, the tablet of ZrSiS was
ground into a fine powder and then sealed in an evacuated quartz ampoule with 5mg/cm$^3$ iodine. The
quartz ampoule was treated in a two-zone tube furnace with a thermal gradient of
about 1100 $^\circ$C - 950 $^\circ$C. After a period of 8 days, single crystals of ZrSiS
were obtained.

STM measurements were carried out in a commercial ultra-high vacuum
system~\cite{Bu2018PRB,zheng2017study,FeiY_O_vacancy,FeiY_Charge_order}.
An electrochemically etched tungsten tip was treated with field emission on a single crystalline of the Au (111) surface.
The samples were cleaved {\it in situ} at liquid nitrogen temperature and immediately inserted into the STM head.
A bias voltage $V_\mathrm{b}$ was applied to the sample, and the tunneling current collected from the tip was
maintained at a fixed setpoint $I_\mathrm{s}$ by a feedback loop control.
All data were acquired at liquid helium temperature ($\sim$~4.5 K).
The differential conductance ($dI/dV$) spectrum was acquired with a standard lock-in technique
with modulation of 10 mV at 1213.7 Hz. The integration time is 3 ms for a single spectrum in 2D measurement.
In a grid measurement (2D or 1D), the tip was moved to a different grid point in constant current mode.
At each grid point, the feedback was turned off while taking the corresponding $dI/dV$ spectrum.
Afterward the feedback was turned on again, and the tip was moved to the next point for data collection.
The DFT calculations were carried out using the Vienna $\it{ab}$ $\it{initio}$ simulation package (VASP)
~\cite{Kresse96efficient, Kresse99paw, Perdew96GGA, Marzari97prb}.
A $1\times1\times5$ supercell with a vacuum layer larger than $2$ nm was
applied in the slab model.

\section{Results and Discussion}
\label{sec3}

\begin{figure}
\centering
\includegraphics[width=0.95\columnwidth]{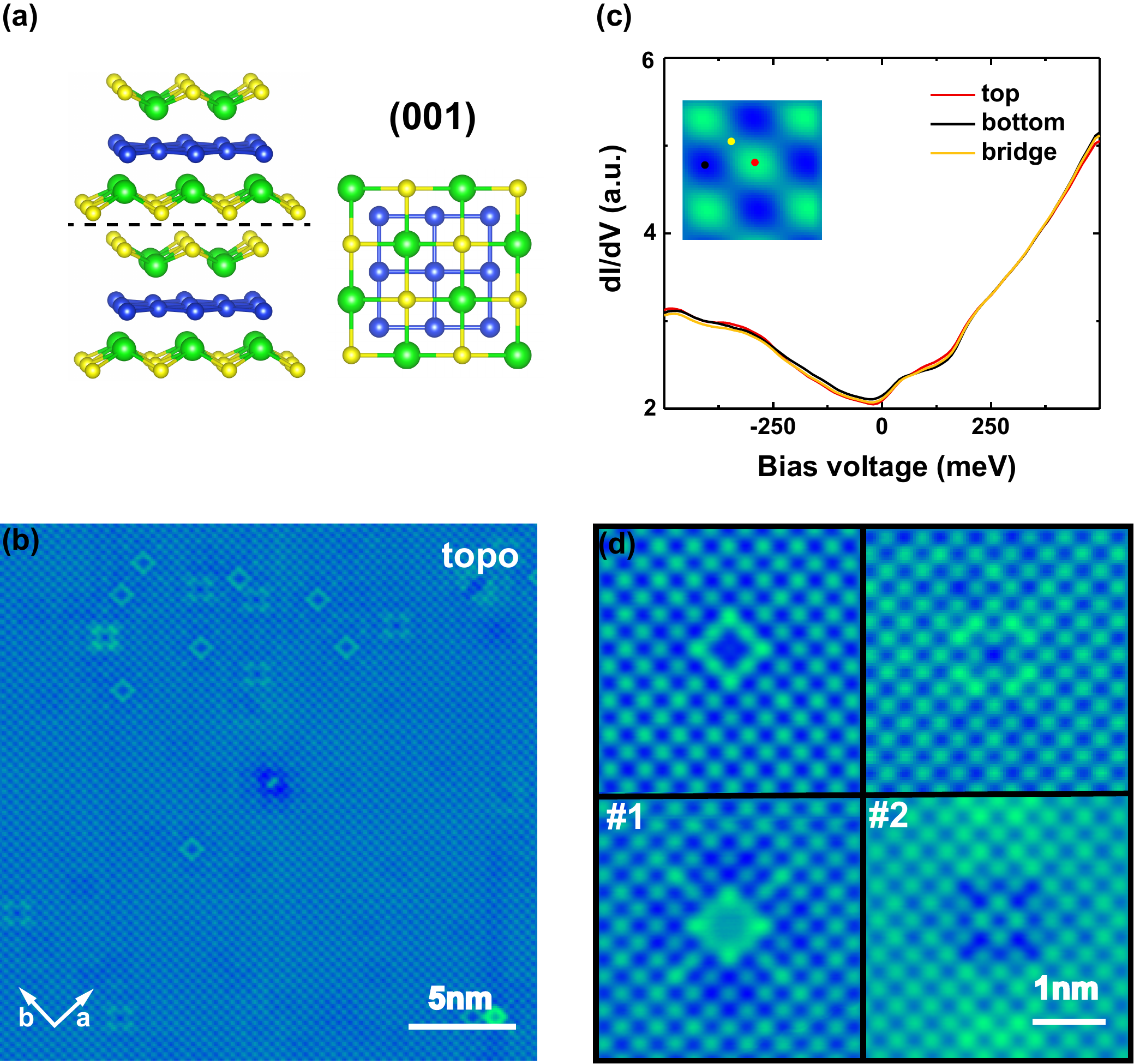}
\caption{(a) Crystal structure of ZrSiS, with a cleavage plane between S layers. Top view of the
cleaved surface is shown on right. The yellow, green and blue dots represent S, Zr and Si atoms, respectively.
(b) A $25\times25$ nm$^2$ topography of ZrSiS taking under $I_s=1$ nA and $V_{\mathrm b}=600$ mV.
The two perpendicular white arrows represent lattice directions.  (c) The average $dI/dV$ spectra
at the top (red), bridge (orange) and hollow (black) sites in the supercell
image (5.2 $\times$ 5.2 {\AA}$^2$), which is shown in the inset.
(d) Four different point defects in topography under the same bias voltage $V_{\mathrm b}=500$ mV.}
\label{fig_n01}
\end{figure}

The family of ZrSiX (X $=$ S, Se, Te) shares a layered crystal structure.
In Fig.~\ref{fig_n01}(a), the crystal structure of ZrSiS shows
that a square lattice of Si atoms is sandwiched between two
sets of ZrS bilayers with glide mirror symmetry~\cite{xu2015two,Wang1995Main}.
Then the crystal structure of ZrSiS is non-symmorphic with Si lattice as
the mirror plane. After an inversion towards the mirror plane and a glide
of the $ab$ plane with a vector of $(1/2,1/2)a_0$ (where $a_0$ is the lattice
constant of the ZrS bilayer), the crystal structure becomes the same as the
original one. This non-symmorphic symmetry is critical to the topological
properties of ZrSiS. For STM experiment, the single crystal sample is
cleaved between two ZrS bilayers, with a S layer naturally exposed to
be the surface plane. Figure~\ref{fig_n01}(b) displays a typical topography
taken on the exposed S surface, with a tunneling junction of
$V_\mathrm{b}=600$ mV and $I_\mathrm{s}=1$ nA. In the topography, a clear square lattice
can be observed with an interatomic spacing of $a_0\approx 3.6$ {\AA}. For the family of ZrSiX, the
density of states (DOS) around the Fermi level is mainly contributed by $d$ electrons of Zr
atoms~\cite{xu2015two}. Top sites of the square lattice are determined to be at
locations of Zr atoms, even though Zr atoms are beneath the cleaved surface plane of S atoms.

Different from that in ZrSiSe~\cite{Bu2018PRB}, our ZrSiS experiment shows a
bias-independent topography, without a shift of square lattice for different
bias-voltage polarities. In a clean area of the sample (12 $\times$ 12 nm$^2$), we performed
a 2D $dI/dV$ spectrum measurement, with the topography acquired simultaneously.
A supercell image was created by overlaying portions of the topography, following
the algorithm in Ref.~\cite{Lawler10nature,Zeljkovic12NatMat}. The supercell image is shown
in the inset of Fig.~\ref{fig_n01}(c), based on which
the measured $dI/dV$ spectra are separately extracted over the top, hollow, and bridge sites.
As shown in Fig.~\ref{fig_n01}(c), the spectra at different sites are almost indistinguishable.
They all exhibit a nonzero DOS around the Fermi level (zero bias), while the
DOS of occupied state is smaller than that of the empty state. The spatially homogeneous
spectrum is consistent with the bias-independent topography.

In Fig.~\ref{fig_n01}(b), sparse point defects of different types can be observed
within the square lattice. A typical diamond- and X-shaped defects are enlarged in
the top row in Fig.~\ref{fig_n01}(d), whose centers are at Zr and S sites, respectively.
In previous STM studies of both ZrSiS and
ZrSiSe~\cite{lodge2017observation,butler2017quasiparticle,Bu2018PRB,ZZ2018NatCom},
the diamond (X-shaped) defects are found to selectively scatter the electronic
surface band (bulk band). For ZrSiSe, a strong scatterer is found to scatter both
the surface and bulk bands~\cite{Bu2018PRB}, which is hitherto not
reported in ZrSrS. In the bottom row of Fig.~\ref{fig_n01}(d), we show two different types
of atomic point defects in ZrSiS, QPI around which is the main focus of this paper.

\begin{figure}[tp]
\includegraphics[width=0.95\columnwidth]{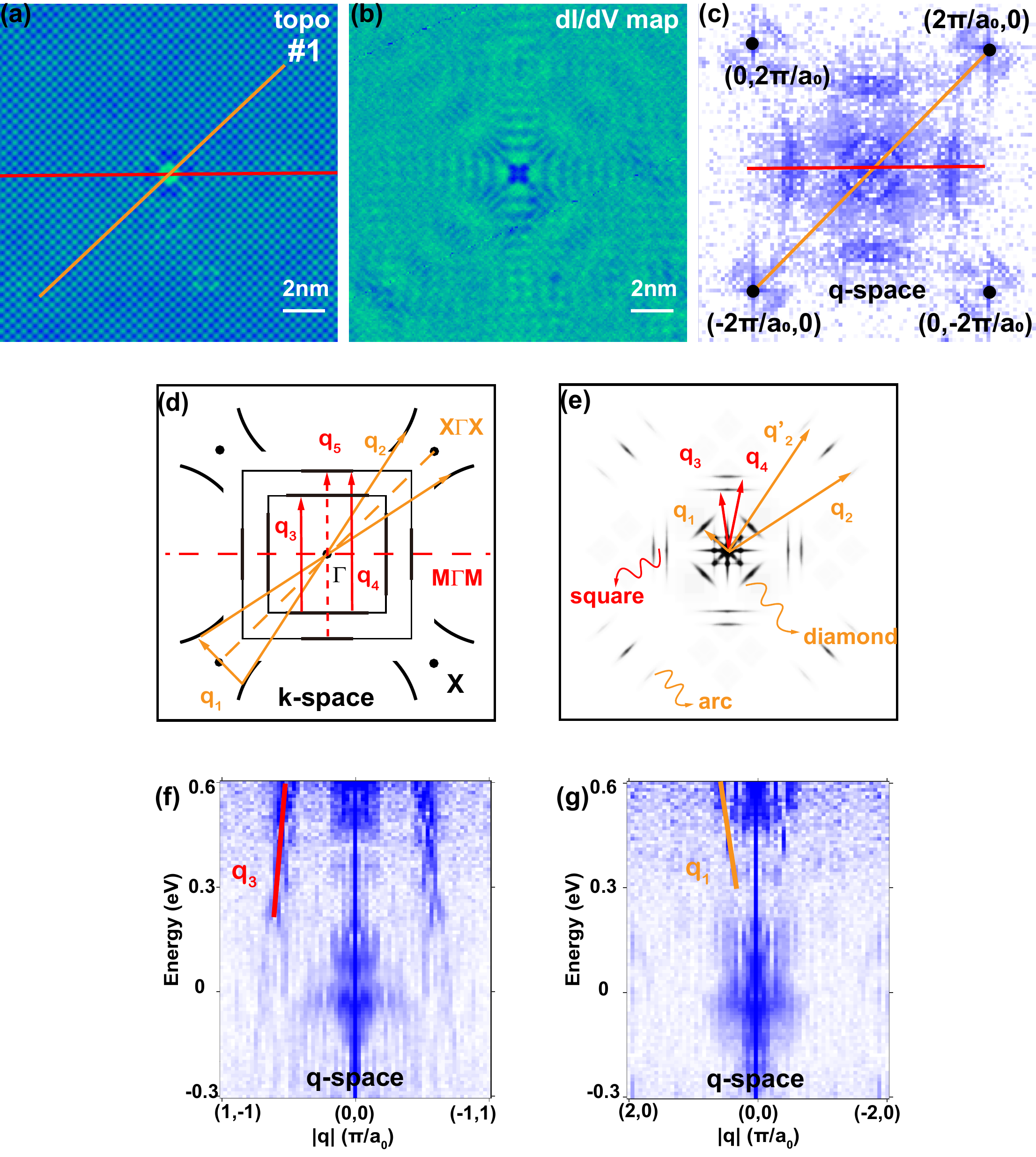}
\caption{(a) A $16\times16$ nm$^2$ topography under  $V_{\rm b} = 300$ mV with
defect $\#$1 at the center. The orange and red line across the defect are along
the lattice and diagonal directions, respectively.
(b) The $dI/dV (V=300~ \mathrm{mV})$ map simultaneously taken with (a).
(c) Fourier transform of the $dI/dV$ map in $\bm q$-space.
(d) A CCE model in $\bm k$-space, and (e) the calculated $\bm q$-space map following this CCE model.
(f-g) QPI energy dispersions along red and orange lines shown in (c).
}
\label{fig_n02}
\end{figure}

The bottom left defect in Fig.~\ref{fig_n01}(d) is centered at the Zr site, around which
a larger topography ($16 \times 16$ nm$^2$ ) is shown in Fig.~\ref{fig_n02}(a).
To study QPI around a single atomic defect, a standard method
is to obtain the $dI/dV$ maps from spectroscopy measurement. For a single
$dI/dV$ map, a quick procedure is to collect the $dI/dV$ signal at the fixed bias voltage
while scanning the tip in constant current mode. Along with the constant-current
topography in Fig.~\ref{fig_n02}(a), a $dI/dV$ map was taken simultaneously at
$V_\mathrm{b}=300$ mV. As shown in Fig.~\ref{fig_n02}(b), this $dI/dV$ map exhibits an
obvious pattern of standing wave centered around the point defect, referred as a QPI
image later. The standing wave originates from the point-defect-induced
scattering between electronic states of different wave vectors
(initial ${\bm k}_i$ and final ${\bm k}_f$) but the same energy. In Fig.~\ref{fig_n02}(b),
the QPI image is not azimuthally symmetric but shows strong oscillations along the
lattice direction and the diagonal ($45^\circ$) direction. Fourier transform of
the $dI/dV$ map is calculated and drawn in Fig.~\ref{fig_n02}(c). Similar to the previous
report for ZrSiSe~\cite{Bu2018PRB}, the Fourier-transformed QPI pattern can be
mainly partitioned into three groups: the central diamond, the concentric
square, and four sets of parallel lines around Bragg peaks.

The QPI pattern is described in the momentum $\bm q$ space with ${\bm q}={\bm k}_f-{\bm k}_i$.
Figure~\ref{fig_n02}(d) shows a contour of constant energy (CCE) model similar to that
in Ref.~\cite{Bu2018PRB}. For the floating surface band~\cite{ZZ2018NatCom, Surface_floating}, there are
four pairs of short parallel arcs around four X points. The scattering between parallel
arcs in the same pair (${\bm q}_1$) corresponds to the central diamond in the QPI pattern.
Scattering between short arcs in diagonal pairs (${\bm q}_2$) corresponds to the
parallel lines around Bragg peaks in the QPI pattern shown
in Fig.~\ref{fig_n02}(e). For the bulk band, two concentric
squares in the CCE model may contribute to concentric squares in the QPI pattern, with
possible wave vectors ${\bm q}_3$, ${\bm q}_4$ and ${\bm q}_5$.
In the QPI pattern in Fig.~\ref{fig_n02}(c), the scattering of both
the surface band and bulk band can be identified, confirming the discovery of a new type of
scatterer in ZrSiS. For the concentric squares in the QPI pattern, either a single square
or two squares have been found for different point defects in
ZrSiX~\cite{lodge2017observation,butler2017quasiparticle,Bu2018PRB,ZZ2018NatCom}.
The concentric square indicated by wave vector ${\bm q}_5$ has
never been found in literature and our experimental results.
Without a high SNR in the QPI pattern, it is hard to judge whether the results are
intrinsic characteristics of the point defect or just vague and indistinct signals
with limited SNR. For simplicity, we intentionally forbid the scattering between
outer to outer squares when calculating the $\bm q$-space map in Fig.~\ref{fig_n02}(e).

To study the energy-dependent QPI pattern, a three-dimensional (3D) dataset has to be taken.
For each spatial point in the scan area, the feedback loop is temporarily interrupted and
a $dI/dV$ spectrum is taken in a selected voltage range, with the energy $E=eV$. After the measurement,
the energy-dependent $dI/dV$ maps can be extracted from the 3D dataset for further analysis.
A long time measurement is necessary for this process (e.g. 12-24 hours), in which the system
instability and thermal-drift affect the SNR. To display the energy-dependent
QPI result, the Fourier-transformed result is often shown along a high symmetry direction in
 $\bm q$ space and plotted as a function of the energy. As shown in Fig.~\ref{fig_n02}(f),
the Fourier-transformed result is displayed along the high symmetry direction,
from $(1,-1){\pi}/{a_0}$ to $(-1,1){\pi}/{a_0}$ in $\bm q$-space
[red line in Fig.~\ref{fig_n02}(c)]. The concentric square
in the QPI pattern intersects with this line at the wave vector ${\bm q}$, later confirmed as ${\bm q}_3$.
In Fig.~\ref{fig_n02}(f), the energy-dependent dispersion of ${\bm q}_3$ can be observed, as
guided by the red solid line. Similarly, figure~\ref{fig_n02}(g)
shows the Fourier-transformed result along the orange line in
Fig.~\ref{fig_n02}(c), from one Bragg peak of $(1,0){2\pi}/{a_0}$ to the diagonal Bragg
peak of $(-1,0){2\pi}/{a_0}$ in $\bm q$-space. The diamond in
the QPI pattern intersects with this line at ${\bm q}_1$, and the dispersion of
${\bm q}_1$ can be observed in Fig.~\ref{fig_n02}(g). We can observe a limited SNR
in the energy-dependent results, which hinders a precise extraction of
dispersions of ${\bm q}_3$ and ${\bm q}_1$ branches.

\begin{figure}[tp]
\includegraphics[width=0.95\columnwidth]{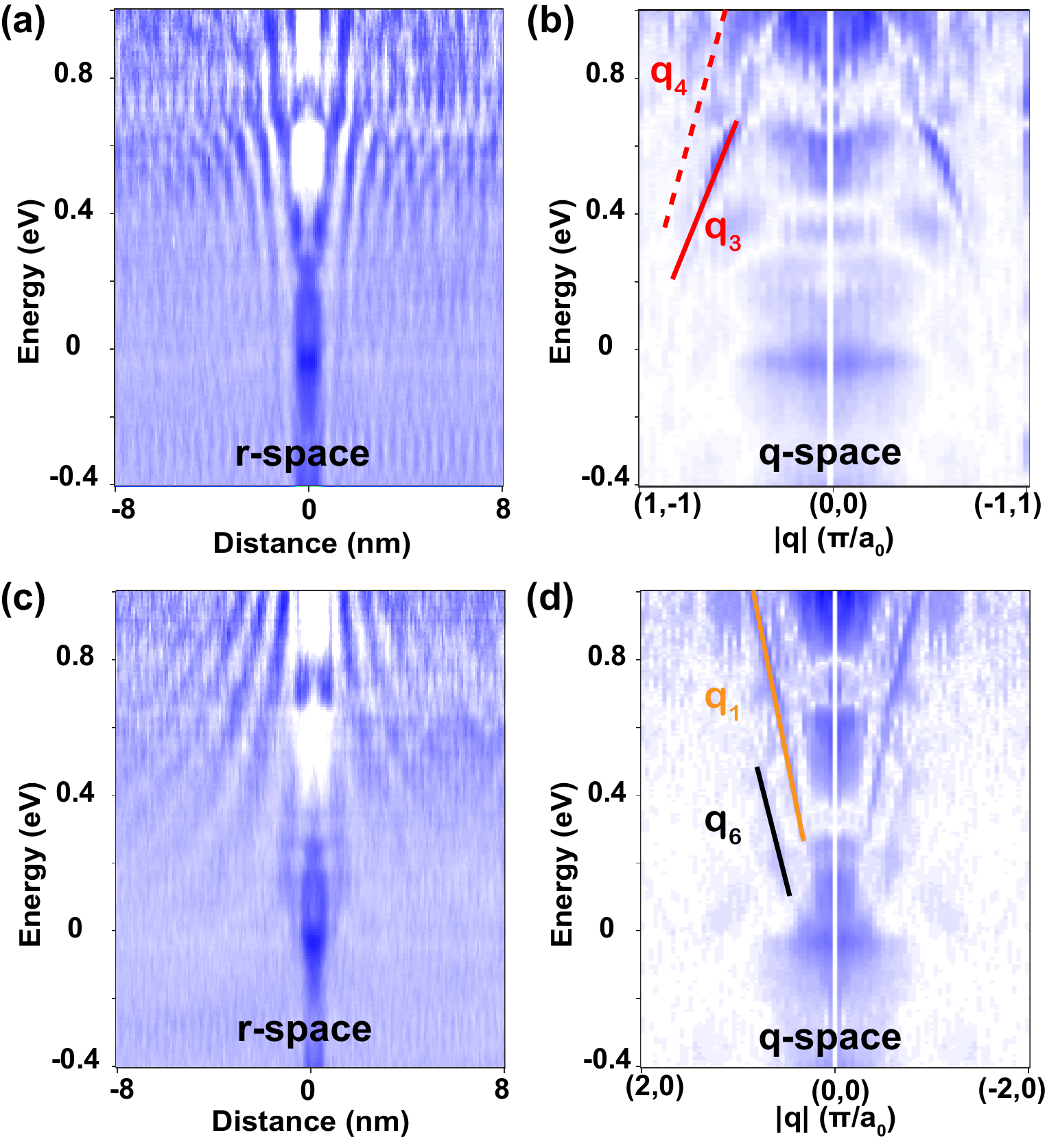}
\caption{(a) Linecut measurement of $dI/dV$ spectrums along the red line in Fig.~\ref{fig_n02}(a) for defect $\#$1.
(b) Fourier transform of the linecut measurement in (a). Maximum of energy dispersion
is guided with a red solid line, and labeled by ${\bm q}_3$. A red dashed line represents the absence
of possible scattering of ${\bm q}_4$ branch.
(c) Linecut measurement of $dI/dV$ spectrums along the orange line in Fig.~\ref{fig_n02}(a) for defect $\#$1.
(d) Fourier transform of the linecut measurement in (c). Maximum of two energy dispersions
are guided with orange (${\bm q}1$) and black (${\bm q}6$) lines, respectively.
 }
\label{fig_n03}
\end{figure}

Here in this work, we introduce a simple and intuitive method, a projective quasiparticle
interference (PQPI) on a single defect, to study the same energy-dependent scattering and extraction
of the electronic band structure. In the two-dimensional QPI image [Fig.~\ref{fig_n02}(a)],
the standing wave propagates strongly along the lattice direction and the diagonal
direction. In this PQPI method, the $dI/dV$ spectra were measured at dense spatial points
along with two corresponding linecuts as labeled in Fig.~\ref{fig_n02}(a). By decreasing
dimension from 2D to 1D in real space measurement, we could increase the average
number in the spectroscopy measurement. In the following 1D linecut measurement, each
spectrum is acquired with the same parameters as in 2D measurement but averaged 5 times.
The effect of system instability and thermal-drift is also lessened within the
short measurement time (e.g. half an hour for a single linecut).

In Fig.~\ref{fig_n03}(a), the measured $dI/dV$ spectrum is shown as a function of
energy (each vertical line), along the linecut of diagonal direction. For each energy,
the oscillating standing wave can be observed along the linecut in the real space.
The real-space signal can be further Fourier-transformed, leading to the $\bm q$-space
QPI pattern along the high symmetry direction from $(1,-1){\pi}/{a_0}$ to $(-1,1){\pi}/{a_0}$.
As shown in Fig.~\ref{fig_n03}(b), the ${\bm q}_3$ branch is more clearly identified,
from the strongly enhanced SNR in the PQPI measurement.
In the meantime, there is no clear dispersion signal of scatter wave vector
${\bm q}_4$ (guided by a red dashed line) in Fig.~\ref{fig_n03}(b).
Because of the short measurement time in PQPI, the energy range is enlarged
to $[-400, 1000]$ meV with an energy resolution of $4$ meV.
A similar spectroscopy measurement was taken
along the lattice direction, with the real-space data shown in Fig.~\ref{fig_n03}(c).
The Fourier-transformed result in $\bm q$-space is shown in Fig.~\ref{fig_n03}(d),
in which the ${\bm q}_1$ branch exhibits a clearly resolved
dispersion (later confirmed by DFT calculations). With the
relative high SNR, another ${\bm q}_6$ branch is also identified,
which will be discussed later. Putting Fig.~\ref{fig_n03}(b) and ~\ref{fig_n03}(d)
together, we conclude that this special impurity scatters both electronic surface
and bulk band, similar to the special defect found in ZrSiSe~\cite{Bu2018PRB}.
Only one ${\bm q}_3$ branch is identified for the scattering within concentric squares.

\begin{figure}[tp]
\includegraphics[width=0.8\columnwidth]{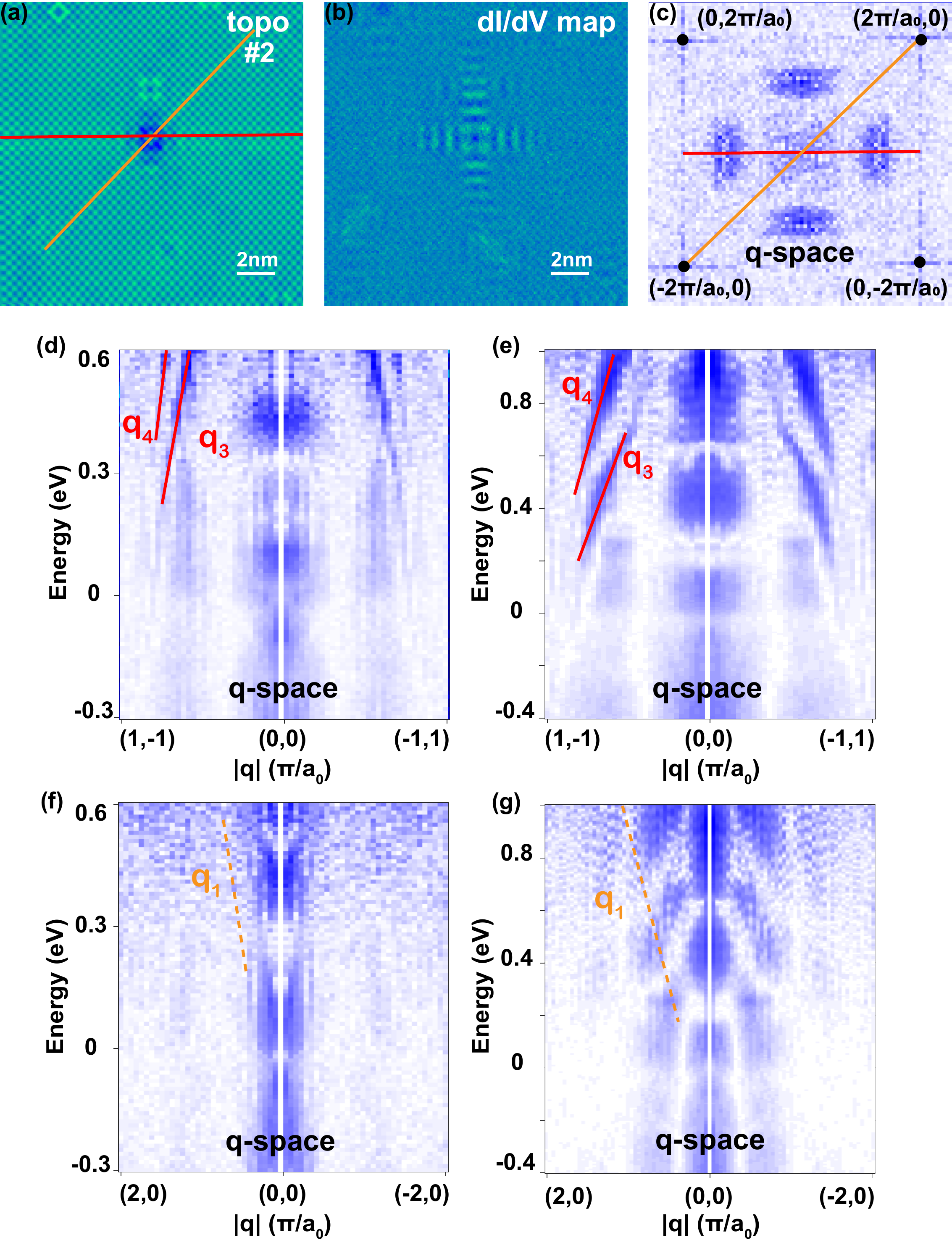}
\caption{
(a) A $16\times16$ nm$^2$ topography under $V_{\rm b} = 600$ mV with defect $\#$2 at the center.
(b) The $dI/dV (V=300~ \mathrm{mV})$ map simultaneously taken with (a).
(c) Fourier transform of $dI/dV$ map in $\bm q$-space.
(d-e) The energy dispersions along the M-$\Gamma$-M direction. The result in (d) is extracted from a
standard 2D $dI/dV$ map along the red line in (c). The result in (e) is extracted from a linecut
measurement along the red line in (a). Two red guide lines highlight the scattering of ${\bm q}_3$ and ${\bm q}_4$ branches.
(f-g) The energy dispersions along the X-$\Gamma$-X direction. The results are extracted from
two different datasets similar to that described in (d-e). An orange dashed line represents
the absence of possible scattering of ${\bm q}_1$ branch.
}
\label{fig_n04}
\end{figure}

Now we turn to the second type point defect ($\#2$). As shown in the bottom right
image in Fig.~\ref{fig_n01}(d), this type of defect is centered at the S site. A larger
topography around this impurity was taken [Fig.~\ref{fig_n04}(a)]. For the same
field of view, the $dI/dV$ map at the bias voltage $300$ mV was also simultaneously taken.
As shown in Fig.~\ref{fig_n04}(b), the standing wave around this defect is observed to
mainly propagate along the $45^\circ$ direction with respect to the lattice direction.
With a nearby X-shaped defect, the QPI image of our interest is
partially affected by the X-shaped-defect-induced standing wave.
The $dI/dV$ map in Fig.~\ref{fig_n04}(b) is Fourier transformed, leading to the QPI pattern
in $\bm q$ space in Fig.~\ref{fig_n04}(c). The concentric squares appear in
the center of the QPI pattern, while the diamond and four sets of parallel lines
around Bragg peaks are absent. This defect seems only scatter the bulk band which is
similar to the X-shaped defect. From the 3D spectroscopy dataset, figure~\ref{fig_n04}(d) shows the
extracted Fourier-transformed result along the $\bm q$-space red line in
Fig.~\ref{fig_n04}(c), as a function of energy. We could roughly distinguish two
dispersed lines, labeled as ${\bm q}_{3}$ and ${\bm q}_{4}$ branches, respectively.

For the PQPI measurement, two real-space lines are chosen in Fig.~\ref{fig_n04}(a)
to be away from the extra standing wave from the X-shaped defect. The $dI/dV$ spectrum
was measured along the line of diagonal direction, whose Fourier
transform is performed and shown in Fig.~\ref{fig_n04}(e). Within a large range
of energy, two dispersed branches (${\bm q}_{3}$ and ${\bm q}_{4}$) can be clearly
identified, confirming the two vague dispersions in Fig.~\ref{fig_n04}(e).
Similarly, a series of $dI/dV$ spectra were measured along the line of
lattice direction, whose Fourier-transformed result is presented
in Fig.~\ref{fig_n04}(g). Different from the QPI pattern in Fig.~\ref{fig_n03},
the ${\bm q}_1$ branch of dispersion is obviously absent, consistent
with that in Fig.~\ref{fig_n04}(c) and \ref{fig_n04}(f).
The high SNR result in Fig.~\ref{fig_n04}(g) confirms that
this defect does not scatter the electronic surface band.

\begin{figure}[tp]
\includegraphics[width=0.95\columnwidth]{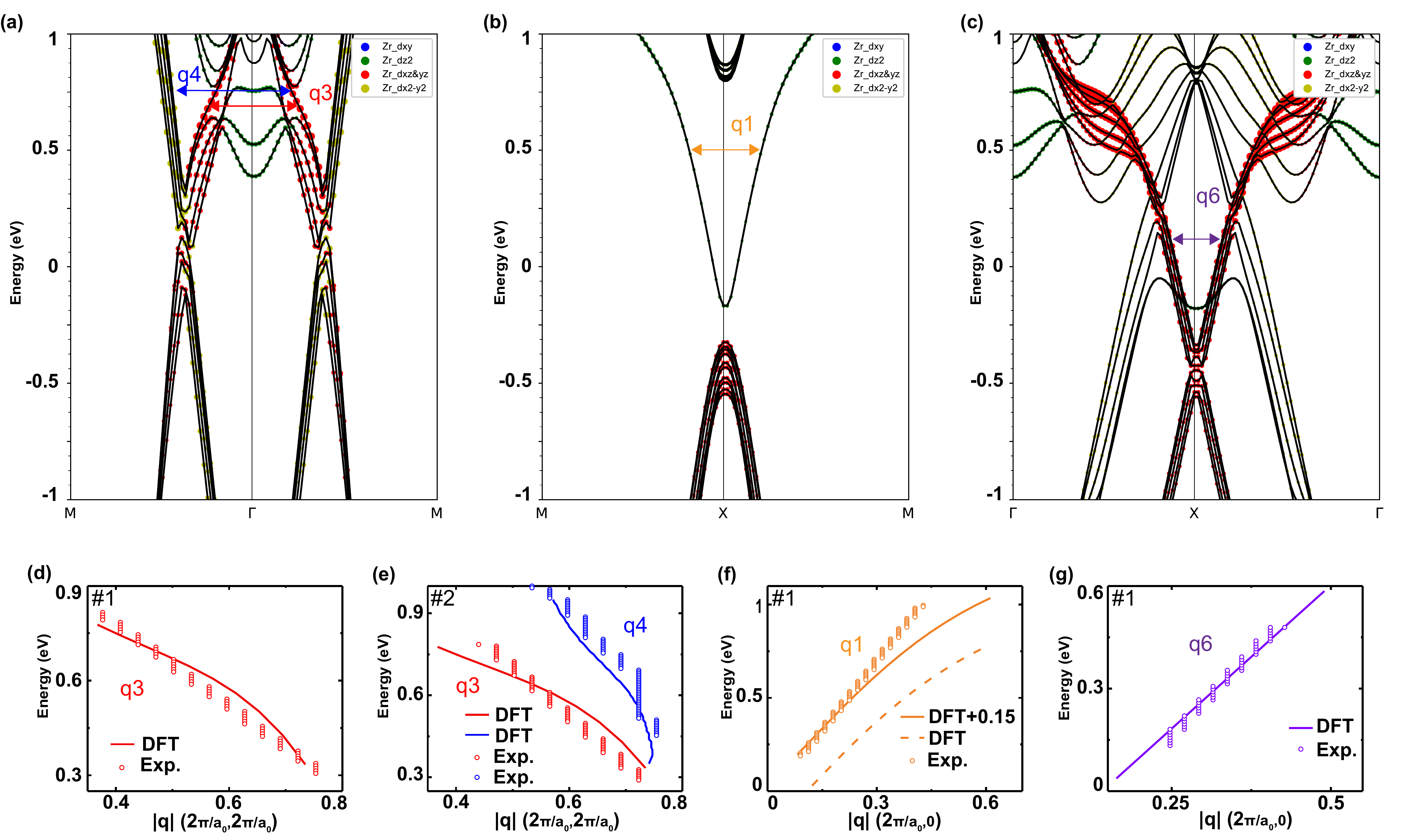}
\caption{(a-c) DFT calculation of the slab band structure along the M-$\Gamma$-M, M-X-M and
 $\Gamma$-X-$\Gamma$ directions in $\bm k$-space. The blue, green, red and orange
 dots represent $d_{xy}$, $d_{z^2}$, $d_{xz}/d_{yz}$ and $d_{x^2-y^2}$ orbitals of Zr atoms, respectively.
(d-e) Comparison of energy dispersions between experimental result
and the DFT calculation along M-$\Gamma$-M direction.
(f-g) Comparison of two energy dispersions between experimental result and
the DFT calculation along X-$\Gamma$-X direction for defect $\#$1. The surface band has
to be shifted $150$ mV up to match with the experimental result in (f).
}
\label{fig_n05}
\end{figure}

With the PQPI method, electronic branches from scattering can be clearly
identified for different defects, which enables a precise extraction of
dispersions and a quantitative analysis of the electronic band structure.
For comparison, the electronic band structure of ZrSiS was calculated
with a DFT of a slab model. Figure~\ref{fig_n05}(a) shows the calculated
band structure along the M-$\Ga$-M direction in $\bm k$-space. The
orbital projection has been considered in the band calculation, as
presented with different colored dots in Fig.~\ref{fig_n05}(a). From
the DFT result, the bands near the Fermi surface are mainly contributed by
different orbital components of Zr atoms. The outer band above the Fermi level
with orange dots is mainly composed of $d_{x^2-y^2}$ components, meanwhile, the
inner branch with red dots is composed of degenerated $d_{xz}$ and $d_{yz}$
components. The QPI branch with wave vector ${\bm q}_3$ corresponds to the
scattering between two internal bands originating from $d_{xz}/d_{yz}$ orbital
of Zr atoms, or two sides of the internal square in the CCE model. The ${\bm q}_4$
branch corresponds to the scattering between one band with $d_{xz}/d_{yz}$ orbital and
another band with $d_{x^2-y^2}$ orbital of Zr atoms. In the CCE model, it is
equivalent to the scattering from one side of the internal square to the opposite
side of the outer square.
The ${\bm q}_5$ branch corresponds to the scattering between two bands with $d_{x^2-y^2}$ orbital of Zr atoms,
which is indicated by a scattering between two sides of the outer square
in the CCE model.

For the defect $\# 1$, we extract ${\bm q} (E)$ from
the dispersed line with high intensity in Fig.~\ref{fig_n03}(b).
We made a constant energy shift of 100 meV for all the bands to present our DFT calculation results.
With this constant shift, the extracted ${\bm q}(E)$
is well consistent with the ${\bm q}_3(E)$ calculated from the electronic band structure [Fig.~\ref{fig_n05}(d)],
which proves that the single branch in Fig.~\ref{fig_n03}(b) matches with ${\bm q}_3$,
instead of ${\bm q}_4$ or ${\bm q}_5$. For the defect $\# 2$, we extract two dispersed branches in Fig.~\ref{fig_n04}(e).
In Fig.~\ref{fig_n05}(e), the two extracted branches are well consistent
with ${\bm q}_3(E)$ and ${\bm q}_4(E)$ calculated from the
electronic band structure. With the decreasing energy, the amplitudes of
${\bm q}_3(E)$ and ${\bm q}_4(E)$ increase, but at different speeds. The dispersion of
${\bm q}_3(E)$ and ${\bm q}_4(E)$ merge around the energy of $300$ meV above the Fermi level,
related with the nodal line in this nodal-line semimetal. The determination of the nodal line
is consistent with the result for ZrSiSe in the previous work~\cite{Bu2018PRB}.

For these two defects, we discover either a single branch of scattering with wave vector ${\bm q}_3$ or
two branches of scattering with wave vectors ${\bm q}_3$ and ${\bm q}_4$. This phenomenon is
similar to that in previous reports for ZrSiS and
ZrSiSe~\cite{butler2017quasiparticle, lodge2017observation, surface_termination}. The
scattering of ${\bm q}_5$ is never discovered heretofore. The high SNR
in our results prove that the absence of scattering of ${\bm q}_5$ is not due to a limited
SNR in the QPI measurement, but an intrinsic property of the scattering. The appearance of
a single branch (${\bm q}_3$) or two branches (${\bm q}_3$ and ${\bm q}_4$) are also clearly
distinguished for two different point defects.

We next extract the scattering between electronic states in the surface band.
Figure~\ref{fig_n05}(b) shows the calculated band structure along the M-$X$-M direction,
perpendicular to the two parallel arcs in the CCE model in $\bm k$-space.
From the calculation, the ${\bm q}_1$ branch happens between bands
with $d_{z^2}$ orbital. This surface band is a floating band, originating
from the surface-induced symmetry breaking from nonsymmorphic
group $P4/nmm$ to symmorphic wallpaper group $P4mm$. With the
broken symmetry, the high band degeneracies are not protected anymore and can be lifted,
resulting in floating or unpinned two-dimensional
surface band~\cite{Surface_floating}. In Fig.~\ref{fig_n05}(f),
we extract ${\bm q}_1 (E)$ from the dispersed line with high intensity
in Fig.~\ref{fig_n03}(d). The main dispersion of ${\bm q}_1$
branch is linear from $300$ meV up to $1$ V. However, the calculated surface band
has to be shifted $150$ meV up to match the experimental results, in addition to the
constant energy shift for all bands. This energy shift of 150 meV
may show the sensitivity of the floating band position with
respect to the impurity~\cite{surface_potential, vacancy-induced}.
The deviation of the calculated surface band above $700$ meV may be related
with a band bending effect in slab model calculation.

In Fig.~\ref{fig_n03}(d), with the high SNR in our PQPI method, a new
branch of ${\bm q}_6 (E)$ can be observed, which has never been reported in previous STM experiments.
The branch of ${\bm q}_6 (E)$ is extracted and shown in Fig.~\ref{fig_n05}(g). After careful comparison,
this branch is found to be consistent with the scattering between bulk bands along $\Gamma X \Gamma$
direction [Fig.~\ref{fig_n05}(c)]. Here the involved bulk bands~\cite{bulk_band}
correspond to the corners of the inner concentric square in Fig.~\ref{fig_n02}(d).
Different from other ${\bm q}$ branches, ${\bm q}_6$ branch is related to a scattering process
between adjacent Brillouin zone (BZ), as illustrated in the Supplementary Material~\cite{spm}.
Normally the inter-BZ scattering is not detectable in QPI patterns. The glide
mirror symmetry however effectively reduces the unit cell by half
and expands the first BZ by two folds. Then the inter-BZ scattering of
${\bm q}_6$ becomes an intra-BZ scattering in the non-symmorphic reshaped
1st BZ, which makes this scattering detectable. A similar picture has been applied to
explain the anomalous half-missing Umklapp feature~\cite{ZZ2018NatCom}.

To explain the complicated defect-dependent scattering is very difficult. The clear scattering
signal, however, enables analysis of symmetry-enforced selection rules~\cite{stern2018prl}.
With preserved non-symmorphic symmetry for the bulk band, the band with
$d_{xz}/d_{yz}$ and $d_{x^2-y^2}$ orbital can be characterized by an integer number
of symmetry flavor $\nu=1$ and $\nu=0$, respectively~\cite{stern2018prl}.
The bands with different $\nu$ induce a direct band crossing and
protected nodal-line (or Dirac ring) in ZrSiX. The bulk band scattering
of ${\bm q}_3(E)$ branch happens between bands with
$d_{xz}/d_{yz}$ orbital, with $\Delta\nu=0$. The scattering of ${\bm q}_4(E)$ branch
happens between bands with $d_{x^2-y^2}$ and $d_{xz}/d_{yz}$ orbital, with $\Delta\nu=1$.
Theoretically, the two branches must be distinguished by matrix elements
because ${\bm q}_3$ corresponds to scattering on the same $\nu$ and ${\bm q}_4$ on
different $\nu$~\cite{stern2018prl}. For defect $\# 1$, only one ${\bm q}_3$ branch is
induced, which means that the orbital character of defect allows the scattering
with $\Delta\nu=0$, while
forbids the scattering with $\Delta\nu=1$. For defect $\# 2$, we see both ${\bm q}_3$
and ${\bm q}_4$ branches imply that the defect does not impose a selection rule, and
the defect should have a mixed orbital character.

In the STM experiment, the tunneling current depends on the overlap between the tip and sample
wavefunctions~\cite{STM_book}. The tip-related effect should also be discussed. The
coexistence of ${\bm q}_3$ and ${\bm q}_4$ implies that the tip also does not impose
a selection rule. The partial overlap between tip and sample wavefunctions is related
to a nonuniversal value of $\beta$, which is defined for the tip in Ref.~\cite{stern2018prl}.
With a vertical $z$-component, $d_{z^2}$ and $d_{xz}/d_{yz}$ orbitals of Zr atoms are prone to be
overlapped with the typical s-wave symmetric tip state~\cite{Spectral_decomposition}.
The overlap between $d_{x^2-y^2}$ orbital (in the $xy$ plane) and the tip state is comparably smaller.
Although with the same $\Delta\nu=0$ as the ${\bm q}_3$ branch, the ${\bm q}_5$ branch
has never been observed. The minimum overlap between $d_{x^2-y^2}$ orbital and the tip state
may lead to a negligible signal of ${\bm q}_5$ in the QPI result. We emphasize that
although the tip-related tunneling is involved in this detection of QPI,
the observed phenomenon is robust against different samples and tips. For example,
there are always two branches of scattering (${\bm q}_3$ and ${\bm q}_4$) for
defect $\# 2$, detected by different tips on different samples. The type of defect is
the key to induce selective scattering of electronic bands.

Although we cannot determine the different defect types yet, further exploration
of the impurity spectrum may provide extra information for a later determination~\cite{Zhou_impurity}.
The most frequently found defects
are the diamond-shaped Zr-site defect and the X-shaped S-site defect, which are expected
to be located within the top ZrS bilayer. When we measure the impurity spectrum for both
defects, no obvious different features can be discerned when compared with the
clean-area spectrum [see Fig. S2(a-b) in the Supplementary Material]. For the Zr-site defect $\# 1$,
in contrast, the central impurity spectrum shows a peak feature around $-40$ meV while
the impurity spectrum at neighboring Zr site shows a peak feature around $350$ meV
[see Fig. S2(c) in the Supplementary Material]. For the S-site defect $\# 2$, instead, the
impurity spectrum at neighboring S site shows a peak feature around $300$ meV
[see Fig. S2(d) in the Supplementary Material]. Although defect $\# 2$ looks like a S-site defect,
we cannot avoid the possibility that it is located at the Zr-site within the bottom ZrS
bilayer. Please note that the neighboring dark S sites and the four radiating dark S lines
share some similarities with that of bright Zr atoms in the topography of
defect $\# 1$. In this possible situation, the Zr-site defects within the top
bilayer are prone to scatter surface band, and the Zr-site defects within the bottom
bilayer only scatter the bulk band. In the future, scanning transmission
electron microscopy (STEM) may be applied to determine the defect types~\cite{Bu_FeSe}.
A comprehensive theoretical model and first-principle calculations are also required to analyze
the orbital character of different defects and explain the defect-dependent scattering
and selection rules.

\section{Summary}
\label{sec4}
PQPI is a complementary tool to analyze the single-defect-induced QPI pattern and its energy dependence.
With a single defect as the scattering center, the QPI oscillation decays with increasing distance away from the defect.
A standard $dI/dV$ mapping of the QPI image is a $2D$ measurement within a small area around the defect.
Although a long time is required in the data-taking procedure, the SNR of the $2D$ measurement is still relatively low.
With an anisotropic propagation of the scattering oscillations,
some high symmetry directions can be chosen in the PQPI method, along which a $1D$ $dI/dV$
measurement can be finished within a short time. Changing from a $2D$ measurement to
a $1D$ measurement, we can increase the data-taking time of every single spectrum
and enhance the SNR of measured results.

In summary, we investigate single-defect-induced QPI oscillations in the nodal-line semimetal ZrSiS.
A new type of Zr-site defect is found to scatter both the bulk band and surface floating band.
With the PQPI method, clear QPI dispersions along high symmetry directions have been clearly resolved.
The clear scattering signal enables a discussion about the non-symmorphic-symmetry-enforced selection
rules. An extra energy shift of the surface floating band is determined and a new branch of
${\bm q}_6$ scattering is discovered. The PQPI method can be generally applied in other
complex materials to explore the distinct interaction between a single atomic defect and electronic states.

\begin{acknowledgments}
We acknowledge and thank R. Queiroz, Z. Fang, X. Dai and H. Weng for discussions and communications.
This work was supported by the National Basic Research Program of China (2019YFA0308602, 2016YFA0300402),
the National Natural Science Foundation of China (NSFC-11374260, 11374261),
and the Fundamental Research Funds for the Central Universities in China.
\end{acknowledgments}


\end{document}


\title{Supplemental Material for \\Projective Quasiparticle Interference of a Single Scatterer to Analyze
the Electronic Band Structure of ZrSiS}

\author{Wenhao Zhang}
\affiliation{Zhejiang Province Key Laboratory of Quantum Technology and Device, Department of Physics, Zhejiang University, Hangzhou, 310027, China}
\author{Kunliang Bu}
\affiliation{Zhejiang Province Key Laboratory of Quantum Technology and Device, Department of Physics, Zhejiang University, Hangzhou, 310027, China}
\author{Fangzhou Ai}
\affiliation{Zhejiang Province Key Laboratory of Quantum Technology and Device, Department of Physics, Zhejiang University, Hangzhou, 310027, China}
\author{Zongxiu Wu}
\affiliation{Zhejiang Province Key Laboratory of Quantum Technology and Device, Department of Physics, Zhejiang University, Hangzhou, 310027, China}
\author{Ying Fei}
\affiliation{Zhejiang Province Key Laboratory of Quantum Technology and Device, Department of Physics, Zhejiang University, Hangzhou, 310027, China}
\author{Yuan Zheng}
\affiliation{Zhejiang Province Key Laboratory of Quantum Technology and Device, Department of Physics, Zhejiang University, Hangzhou, 310027, China}
\author{Jianhua Du}
\affiliation{Zhejiang Province Key Laboratory of Quantum Technology and Device, Department of Physics, Zhejiang University, Hangzhou, 310027, China}
\author{Minghu Fang}
\affiliation{Zhejiang Province Key Laboratory of Quantum Technology and Device, Department of Physics, Zhejiang University, Hangzhou, 310027, China}
\affiliation{Collaborative Innovation Center of Advanced Microstructures, Nanjing University, Nanjing 210093, China}
\author{Yi Yin}
\email{yiyin@zju.edu.cn}
\affiliation{Zhejiang Province Key Laboratory of Quantum Technology and Device, Department of Physics, Zhejiang University, Hangzhou, 310027, China}
\affiliation{Collaborative Innovation Center of Advanced Microstructures, Nanjing University, Nanjing 210093, China}

\maketitle
\makeatletter
\renewcommand{\thefigure}{S\@arabic\c@figure}
\renewcommand{\theequation}{S\@arabic\c@equation}


\begin{figure}
\centering
\includegraphics[width=0.95\columnwidth]{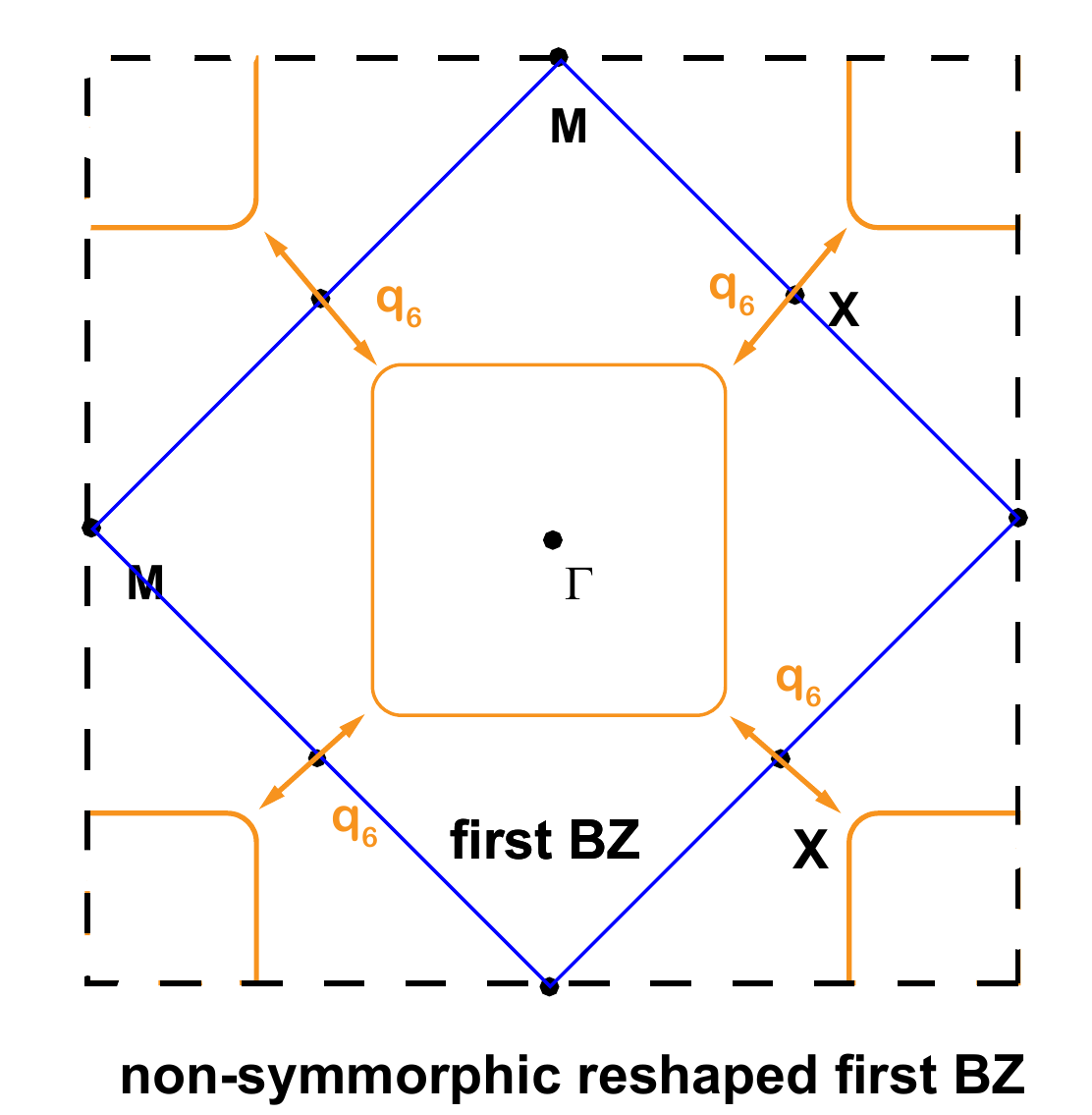}
\caption{Illustration of the ${\bm q}_6$ scattering process. The yellow square represents the bulk band shown as the inner concentric 
square in Fig. 2(b). In this schematic diagram, the scattering process occurs between corners of the bulk band in adjacent Brillouin zone (BZ).  
The glide mirror symmetry effectively reduces the unit cell by half and expands the first BZ (the blue solid-line frame) 
by two folds (the black dash-line frame). The inter-BZ scattering of ${\bm q}_6$ becomes an intra-BZ scattering in the 
non-symmorphic reshaped 1st BZ, which makes this scattering detectable.}
\label{fig_n1} 
\end{figure}

\begin{figure}
\centering
\includegraphics[width=0.95\columnwidth]{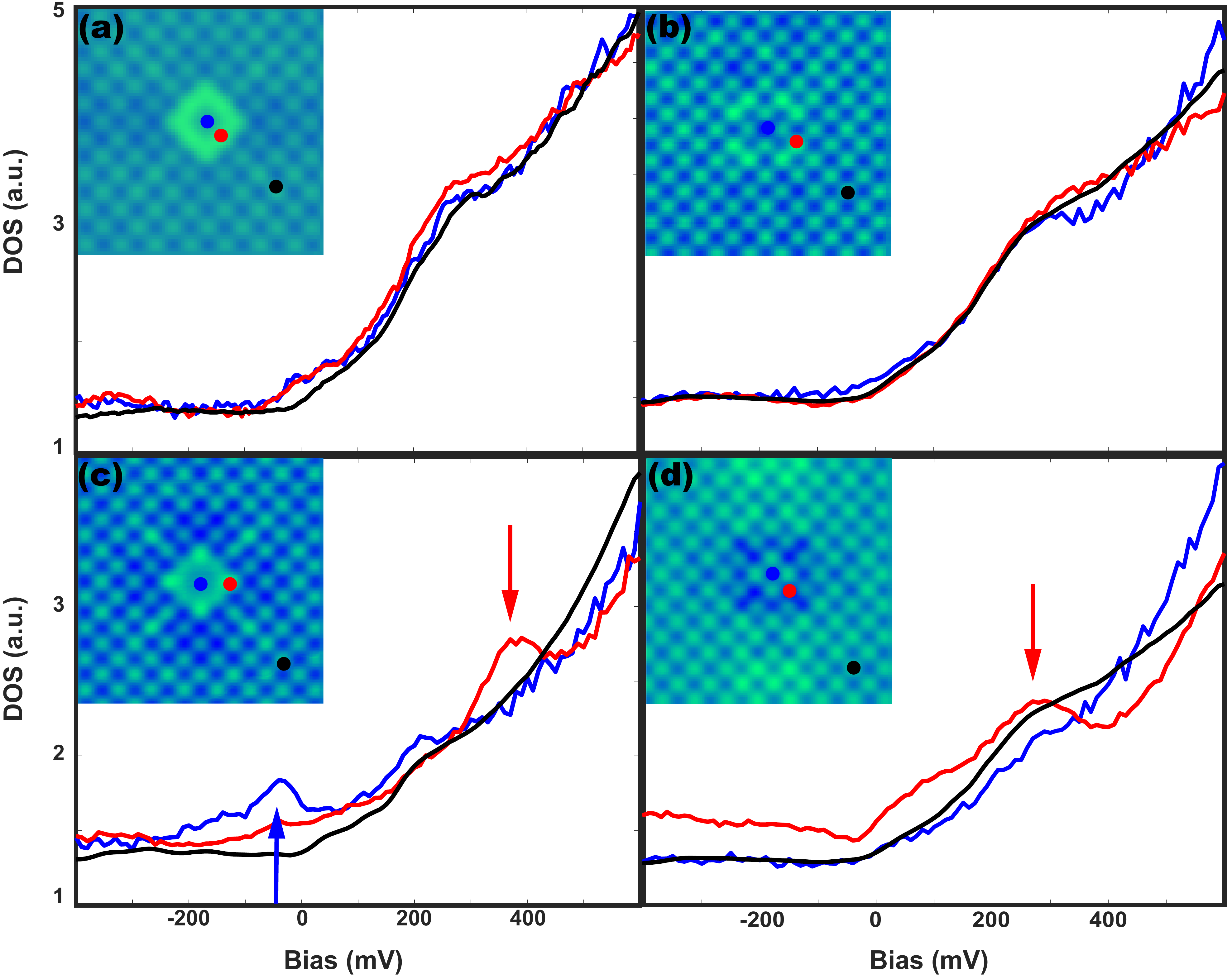}
\caption{The impurity spectrum taken on different types of defects. (a-b) Common Zr and S site defects. The typical spectra show 
no obvious different features compared with the clean-area spectrum. (c) A new type of Zr site defect $\# 1$. The central spectrum shows a peak 
around $-40$ meV and the spectrum taken on neighboring Zr-Site shows a peak around $350$ meV. (d) Uncommon S-site defect $\# 2$.
The spectrum taken on neighboring S-site shows a peak around $300$ meV}
\label{fig_n2}
\end{figure}
